\documentclass[preprint,12pt,authoryear]{elsarticle}

\usepackage{amssymb}
\usepackage{amsmath}

\usepackage{subfig}
\usepackage{enumitem}
\usepackage{graphicx}  
\usepackage{rotating}
\usepackage{multirow}
\usepackage{makecell}
\usepackage{pifont}
\usepackage[normalem]{ulem}
\usepackage{hyperref}
\usepackage{booktabs}
\usepackage{float}
\usepackage{listings}
\usepackage{array} 
\usepackage[table]{xcolor}

\definecolor{lightgray}{gray}{0.9}
\journal{Artificial Intelligence in Medicine}

\begin{document}

\begin{frontmatter}

\title{Modeling Challenging Patient Interactions: LLMs for Medical Communication Training \\ 
\textit{\small A Preprint}}

\author[1]{Anna Bodonhelyi\corref{cor1}}
\author[2]{Christian Stegemann-Philipps}
\author[2]{Alessandra Sonanini}
\author[2]{Lea Herschbach}
\author[3,4]{Marton Szep}
\author[2]{Anne Herrmann-Werner}
\author[2]{Teresa Festl-Wietek}
\author[1]{Enkelejda Kasneci}
\author[2]{Friederike Holderried}

\affiliation[1]{organization={Chair for Human-Centered Technologies for Learning, Technical University of Munich},
            addressline={Arcisstr. 21.}, 
            city={Munich},
            postcode={80333}, 
            country={Germany}}
\affiliation[2]{organization={Tübingen Institute for Medical Education, Eberhard Karls University},
            addressline={Elfriede-Aulhorn-Str. 10.}, 
            city={Tübingen},
            postcode={72076}, 
            country={Germany}}
\affiliation[3]{organization={Department of Orthopaedics and Sports Orthopaedics, TUM University Hospital},
            addressline={Ismaninger Str. 22}, 
            city={Munich},
            postcode={81675}, 
            country={Germany}}
\affiliation[4]{organization={Chair for AI in Healthcare and Medicine, Technical University of Munich and TUM University Hospital},
            addressline={Ismaninger Str. 22}, 
            city={Munich},
            postcode={81675}, 
            country={Germany}}

\cortext[cor1]{Corresponding author: Anna Bodonhelyi, anna.bodonhelyi@tum.de}

\begin{abstract}
Effective patient communication is pivotal in healthcare, yet traditional medical training often lacks exposure to diverse, challenging interpersonal dynamics. 
To bridge this gap, this study proposes the use of Large Language Models (LLMs) to simulate authentic patient communication styles, specifically the ``accuser'' and ``rationalizer'' personas derived from the Satir model, while also ensuring multilingual applicability to accommodate diverse cultural contexts and enhance accessibility for medical professionals. 
Leveraging advanced prompt engineering, including behavioral prompts, author’s notes, and stubbornness mechanisms, we developed virtual patients (VPs) that embody nuanced emotional and conversational traits. 
Medical professionals evaluated these VPs, rating their authenticity (accuser: $3.8 \pm 1.0$; rationalizer: $3.7 \pm 0.8$ on a 5-point Likert scale (from one to five)) and correctly identifying their styles. 
Emotion analysis revealed distinct profiles: the accuser exhibited pain, anger, and distress, while the rationalizer displayed contemplation and calmness, aligning with predefined, detailed patient description including medical history. 
Sentiment scores (on a scale from zero to nine) further validated these differences in the communication styles, with the accuser adopting negative ($3.1 \pm 0.6$) and the rationalizer more neutral ($4.0 \pm 0.4$) tone. 
These results underscore LLMs' capability to replicate complex communication styles, offering transformative potential for medical education. 
This approach equips trainees to navigate challenging clinical scenarios by providing realistic, adaptable patient interactions, enhancing empathy and diagnostic acumen. 
Our findings advocate for AI-driven tools as scalable, cost-effective solutions to cultivate nuanced communication skills, setting a foundation for future innovations in healthcare training.

\end{abstract}

\begin{keyword}
virtual patient \sep large language models \sep medical education \sep conversational agent \sep communication style
\end{keyword}

\end{frontmatter}

\section*{Acknowledgement}
We would like to express our gratitude to Hannes Burrichter for his valuable contributions in developing the communication styles and implementing the framework for the user study.

\section{Introduction}
Simulated medical interactions are a cornerstone of medical education, providing students with a safe environment to hone communication and diagnostic skills \citep{de2009endpoints}. Virtual Patients (VPs) facilitate realistic clinical practice without relying on standardized patients \citep{holderried2024generative, li2024leveraging, grevisse2024raspatient}. Advances in Large Language Models (LLMs) have transformed this field, enabling conversational agents with nuanced and adaptive responses~\citep{rudolph2024ai}. While LLM-based VPs improve accessibility, reduce costs, and enrich medical training, most existing implementations prioritize clinical accuracy or general dialogue coherence, neglecting the replication of diverse communication styles and emotional dynamics essential for authentic patient interactions~\citep{li2024leveraging, grevisse2024raspatient}. However, authenticity alone is not sufficient; VPs must simultaneously create realistic interactions while posing complex, scenario-specific difficulties that actively engage students. 

Artificial Intelligence (AI) medical chatbots have been applied to healthcare and education, supporting tasks like mental health counseling \citep{qiu2024interactive} and history-taking training \citep{holderried2024generative}. Studies highlight the plausibility of LLM-driven chatbots, while tools like CureFun~\citep{li2024leveraging} and RasPatient Pi~\citep{grevisse2024raspatient} showcase their adaptability for cost-effective, customizable VP simulations. Such tools typically rely on illness scripts, which provide detailed descriptions of a patient’s medical history, personal life, and general background to guide realistic and contextually accurate interactions. 
However, existing systems overlook diverse, emotionally charged communication styles, which are crucial for preparing medical trainees for real-world interpersonal dynamics.

This study investigates the potential of LLM-powered VPs to simulate distinct communication styles while being medically accurate, specifically focusing on the accuser and rationalizer personas, as informed by the Satir model \citep{satir}. By leveraging user feedback from psychotherapy professionals at an international conference on psychotherapy, we evaluated the realism and effectiveness of these personas in simulated medical interactions. Furthermore, we explored the use of emotion and sentiment analysis, techniques that detect and classify emotional states and attitudes in text, as tools for quality control, aiming to enhance the performance and reliability of LLM-driven VPs. This work extends the current understanding of AI's role in medical education by highlighting the importance of communication style diversity, offering insights into their implementation, and addressing the limitations of existing approaches in this evolving field. The key research questions explored in this work are as follows: 

\begin{enumerate}
    \item What different prompt engineering techniques can be utilized to develop distinct communication styles, as defined by the Satir~\citep{satir} model, in VP simulations?
    \item How do medical professionals evaluate the realism and adherence to predefined communication styles in LLM-simulated interactions, through questionnaire feedback?
    \item How well do predefined illness scripts of communication styles align with emotion and sentiment analysis metrics in VP interactions?
\end{enumerate}

\section{Related Work}
This section reviews foundational research supporting our study on AI-driven VPs. 
We first introduce the most common patient communication styles, then, we analyze methods for modeling communication styles in conversational AI to simulate realistic dialogue dynamics. 
Lastly, we explore chatbot applications in medical education, emphasizing their role in training students for complex scenarios.

\subsection{Patient Communication Styles}
To model patient communication styles, we adopted the Satir model~\citep{satir}, which categorizes interpersonal communication into five distinct types: appeaser, accuser, rationalizer, distractor, and congruent type. These communication styles specifically describe behaviors exhibited under stress, such as during conflicts or challenging situations like a debilitating illness. 
Each style reflects particular traits and response patterns, forming a structured basis for simulating diverse patient personalities in scenarios characterized by emotional distress. 
The appeaser prioritizes harmony, often suppressing their opinions and struggling with decisions, while the accuser is assertive and confrontational, masking insecurity with dominance. 
The rationalizer takes a detached, logical stance, using monotone speech and extended reasoning to project intellectual authority. The distractor’s spontaneous, fragmented communication shifts attention away from issues, impeding problem-solving. In contrast, the congruent type exemplifies balanced, transparent communication, fostering mutual respect and constructive dialogue. 
Previous research highlights that VP simulations can influence attitudinal outcomes such as confidence, preparedness, and self-efficacy in various clinical scenarios, emphasizing the need for exposure to diverse patient interactions, including challenging personalities~\citep{kononowicz2019virtual}. 
In this work, we focus on modeling the non-cooperative communication styles,  accuser and rationalizer, which are the two most challenging personalities to simulate \citep{sicilia2022modeling}. 
Their traits were carefully aligned with the descriptions in the Satir model~\citep{satir}, ensuring realistic and consistent communication patterns. 
By incorporating both accuser and rationalizer VPs, we simulate distinct conversational challenges that require users to adapt their questioning strategies, regulate emotions, and employ de-escalation techniques, ultimately improving their ability to manage complex real-world patient interactions. 
The communication-type guidelines defined by Satir~\citep{satir} provided a foundational framework for defining how the VPs should interact, ensuring consistency in their responses and serving as a crucial reference for selecting and refining the most effective prompt engineering techniques.

\subsection{Modeling Communication Styles in Conversational AI}
Communication styles consist of the way individuals convey and interpret messages verbally and paraverbally, shaped by traits like dominance, animation, attentiveness, and openness, ultimately influencing the effectiveness of communication \citep{norton1978foundation}. 
It is also influenced by underlying personality traits, the dynamics of interpersonal relationships, and the specific context in which the interaction occurs~\citep{mattar2013strangers}.
They act as a filter that shapes the delivery of information but do not necessarily dictate emotional responses or content. 
The Satir model~\citep{satir}, however, defines communication styles specifically in the context of stress and emotional coping mechanisms. 
Unlike general styles, Satir’s archetypes characterize both how a person speaks and what they say in response to stress, shaping their emotional expression and behavioral patterns. This distinction is essential when designing VP, as it ensures that both verbal delivery and emotional reactivity align with the intended archetype.

Modeling communication styles in conversational AI aims to replicate the unique ways individuals express themselves, such as tone, word choice, and sentence structure, in written interactions. This allows AI systems to effectively take on roles or impersonate fictional personas with realistic behavior, produce natural, context-aware responses, and making written conversations feel authentic and personalized. 
\citet{ruane2021user} show that personality traits can be effectively conveyed through text, enabling users to recognize intended characteristics even without visual or auditory elements. They also emphasize the importance of chatbot personality in shaping user experience, underscoring its value in creating contextually engaging and effective conversational agents.

Conversational AI systems can be designed to adapt to individual communication styles through various methodologies, such as language modeling, style transfer, and representation learning. These techniques, for example, enable the transformation of robotic conversations into human-like, personalized interactions while preserving the original content of the responses~\citep{srivastava2022neural}. \citet{poivet2023influence} show that communication styles, such as aggressive or cooperative approaches, influence user behavior, preferences, and perceptions of conversational agents. In narrative interactions, users form expectations about an agent's communication style based on stereotypes linked to its explicit role, such as associating an opponent with hostility and an aggressive tone~\citep{infante1995teaching, poivet2023influence}. Additionally, adopting social-oriented styles or mimicking users' verbal behavior significantly enhances engagement and rapport, particularly in hedonic interactions~\citep{van2023effects}. Our work builds on these findings by aiming to simulate distinct and contextually appropriate communication styles for medical VPs, ensuring realistic and engaging interactions that align with user expectations.

Our approach differs from previous studies by prioritizing the authenticity of personality representation over user engagement, as our VPs are designed for educational use. We focus on accurately portraying distinct communication styles tied to specific patient personalities, ensuring realism in medical interactions. To assess this, we gather feedback from medical professionals through structured evaluations, examining the realism, adherence to predefined communication styles, and overall effectiveness of LLM-simulated interactions, and also evaluate the conversations with automated emotion analysis. In our work, we focused on prompt engineering techniques, as fine-tuning an LLM is costly, resource-intensive, and requires extensive domain-specific data. While fine-tuning allows for deeper model adaptation, prompting offers a flexible and data-efficient alternative, balancing realism and feasibility in VP simulations \citep{petroni2019language, szep2024practical}. This emphasis on scenario-based personality traits within clinical contexts distinguishes our work from broader studies on personality simulation or general AI-driven communication.

\subsection{Chatbots in Medical Education}
VP tools have become invaluable in medical education, offering significant advantages over traditional methods. They provide a controlled, repeatable environment where students can practice medical interactions without the logistical and financial demands of live patient simulations~\citep{holderried2024generative}. Unlike in-person training, VP tools require no physical spaces or standardized patients, making them cost-effective and widely accessible. They expose students to a broad spectrum of scenarios, including rare conditions and challenging patient interactions, while allowing repeated practice to build skills without real-life risks. Effective communication is essential for gathering information, fostering patient relationships, and supporting clinical decision-making, whereas poor communication can compromise diagnostic accuracy and jeopardize patient safety \citep{de2009endpoints, street2009does, stewart1995effective, goold1999doctor, hausberg2012enhancing}. These underscore the importance of VP tools in honing these critical skills, also granting flexibility and cost-effectiveness for modern medical training.

Recent advancements have sought to enhance their realism and adaptability of VPs by incorporating expert-driven design principles and tailored feedback mechanisms. Roleplay-doh~\citep{louie2024roleplay} introduces an innovative method for creating customized AI patient simulations by translating expert feedback into adherence principles, achieving a 30\% improvement in simulation quality, though it remains limited by its focus on novice counseling scenarios. Similarly, the CureFun~\citep{li2024leveraging} framework uses LLMs to develop accessible, cost-effective VPs for medical education, supporting natural student-simulated patient interactions and providing clinical skill-enhancing feedback, yet it faces challenges in fully assessing diagnostic capabilities within these LLM-based simulations. Another study~\citep{holderried2024generative} shows that GPT-driven chatbots effectively simulate patient interactions for history-taking practice by providing predominantly plausible, script-aligned responses, though occasional responses prioritize social desirability over medical accuracy, highlighting a need for enhanced answer accuracy controls. RasPatient Pi~\citep{grevisse2024raspatient} further explores the possibilities of LLM-based VP tools by integrating automatic speech recognition, LLM, text-to-speech, and avatar visualization to create a customizable, low-cost medical simulation platform. A separate framework~\citep{qiu2024interactive} uses two LLMs in role-play to simulate counselor-client interactions, demonstrating the potential of LLMs in psychological counseling training. Lastly, the VirCo~\citep{rudolph2024ai} chatbot can generate realistic responses that match the flow of conversation while maintaining a consistent identity. However, differences in ratings between human evaluators and inconsistencies with GPT-3.5 and GPT-4 — especially GPT-3.5 — make it challenging to accurately assess the chatbot’s coherence. These tools offer valuable opportunities for medical students to practice, yet a gap remains in understanding how effectively LLMs can portray specific challenging personalities. While VPs can reliably stay in character and deliver necessary medical information, their ability to authentically reflect distinct personalities remains uncertain, especially in challenging scenarios involving stress, pain, or emotional distress. Since LLMs are inherently designed to generate responses that align with user expectations, they may default to agreeable or overly accommodating tones~\citep{goetz2023unreliable, chen2024don}. In this work, we aim to address this and provide medical students with a more realistic setting to experiment with diverse communication strategies and enhance their ability to communicate effectively.

\section{Methodology}
In this work, we develop two distinct VP roles based on the Satir model \citep{satir}: the accuser and rationalizer types. In this section, we detail the development and implementation of the selected communication types within the framework.

\subsection{Framework}
We implement our VP communication styles using the framework proposed by \citet{holderried2024generative}, an LLM-driven tool designed to support clinical skill development, as shown in~\autoref{fig:structure}. While the tool typically provides feedback to students on medical history-taking, we modified the system for our data collection phase by substituting this feedback with a questionnaire tailored to gather insights from our study participants instead of giving feedback on medical history coverage. The chat interface and the creation of VPs is based on a role-playing approach~\citep{holderried2024generative, louie2024roleplay, li2024leveraging}.
Participants can engage with the VPs in English and German, ensuring accessibility, since psychotherapy professionals prefer to use their native language, in general \citet{sercu2023psychotherapy}.

\subsection{Technical details}
Similarly to the original framework~\citep{holderried2024generative}, the tool facilitates interactions between users and a GPT-based VP through a chat interface that allows students to pose questions and receive responses. This interface uses Django and JavaScript for processing and TailwindCSS for layout. Communication with GPT is enabled via OpenAI’s API, using openAI's python sdk to make requests to the chat/completions endpoint with GPT-4.

Expanding on the framework, we enhanced its functionality by incorporating communication styles derived from the Satir model~\citep{satir}. These predefined patterns of interaction, tailored to challenging scenarios, were adapted specifically for use in VP simulations, enriching their realism and behavioral diversity.

\subsection{Prompt Engineering}

\paragraph{\textbf{Author's Note and First Message}}
To enhance the realism and coherence of the role-play, we employed two key techniques: an author's note and a first message. The author's note functions as a dynamic system message that continuously reinforces the VP’s personality, tone, and behavior, ensuring alignment with its intended archetype~\citep{sillytavern2024}. This mechanism acts as a filter, guiding how the VP communicates. Simultaneously, we specify the first assistant message to establish the mood and context of the interaction, along with setting the stage for accurate role-playing~\citep{kong2023better, sillytavern2024}. To further enhance authenticity, we incorporated hidden thoughts and feelings visible only to the LLM, promoting consistent and nuanced responses while preserving the natural flow of conversation (\autoref{lst:first_message}).

\paragraph{\textbf{Behavior Instruction Prompts}}
To enhance the system, we further refined the prompt design by incorporating a chatbot-optimized illness script and a behavioral instruction prompt with dedicated fields (Section \ref{chap:behav_prompt}). These fields are tailored for each VP and include coded categories such as patient details, medical concerns, accompanying symptoms, medical and family history, which we extended with communication style specific entries. To achieve consistency in the VPs' communication styles, we modified the framework~\citep{holderried2024generative} and employed additional fields to provide more granular sociocultural information about the patient, including mood, communication tendencies, and typical reactions. When designing the prompts, we carefully considered potential user intents \citep{bodonhelyi2024user, bhatt2024med, narendran2024healthbot} within the broader use case of medical students conducting history-taking, ensuring that the AI-generated responses aligned with realistic clinical interactions. These refinements allowed for more precise and structured guidance, ensuring the VPs maintained realistic and contextually appropriate behaviors throughout the interactions. While the initial model emphasized maintaining the LLM's role fidelity and delivering medically accurate information based on patient records, our extended framework ensures dynamic, context-sensitive interactions aligned with the VP's communication style, creating a challenging conversational situation between the users and the VPs.

\paragraph{\textbf{Stubbornness Mechanism}}
To further promote challenging interactions, specific communication style features were incorporated into the framework, ensuring that both VPs consistently declined any association between their symptoms and psychological factors. They resisted discussing family issues and expressed skepticism toward therapy, adding an additional layer of complexity to the interactions. These refinements make the VPs' behavior more realistic and provide a nuanced platform for training handling difficult conversations. One possible solution would be to use a separate LLM to monitor the conversation and ensure that the need for persuasion is addressed. However, implementing this approach lies outside the scope of this study. To ensure the VPs followed a strict if-else structure and only responded positively to psychological solutions when specific conditions were met, we provided example sentences. These examples help guide the LLM, reinforcing the intended behavior in relevant scenarios (\autoref{lst:stubborn}).

\section{Procedure of the User Study}
\begin{figure*}[t]
	\centering
	{\includegraphics[height=5.2cm, keepaspectratio]{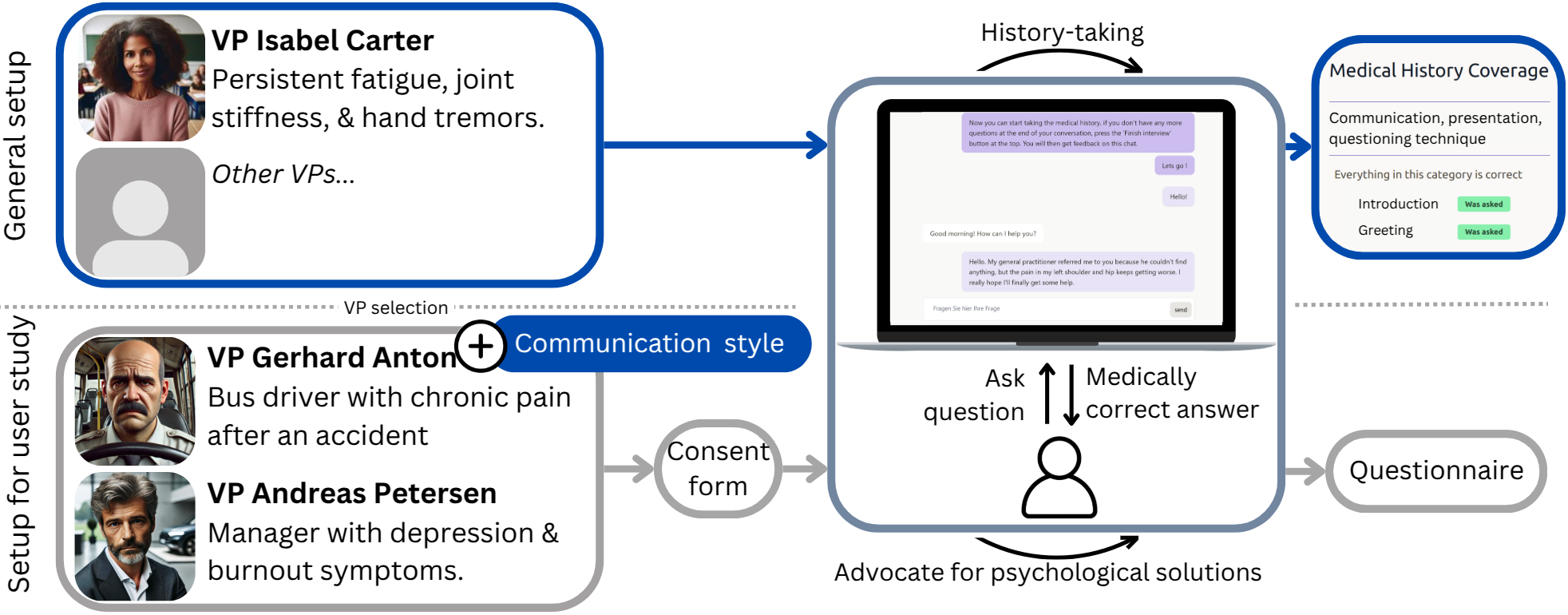}}
	\caption{User study setup compared to the general tool usage. The VP Isabel Carter, is a fictional character created solely for demonstration purposes.}
    \label{fig:structure}
\end{figure*}

For the purpose of evaluating the authenticity of the VPs, we modified the original framework proposed by \citet{holderried2024generative} to collect psychotherapy professional feedback instead of providing feedback to students on medical record-taking (\autoref{fig:structure}). This is done with a questionnaire tailored to gather insights from our study participants, focusing on communication style consistency and the educational value of the simulated interactions. We established a fixed environment~\citep{bodonhelyi2024passive} to reduce variability and maintain control during evaluation, ensuring a consistent design for both VPs, where users could interact with each similarly. Participants had the option to choose between two languages for the VP and questionnaire: English and German. Following agreement to a consent form for data collection, participants were presented with a brief introduction of the VP. The VPs were assigned randomly to ensure balanced representation across conditions, maintaining equal distribution among participants. They then engaged in open-ended conversation with the respective VP, representing the distinct communication types. Once participants felt their interaction was complete, they proceeded to complete a questionnaire to evaluate the experience.

\subsection{Study Design}
The study was conducted during an international conference on psychotherapy, namely ICPM 2024 (27th World Congress of the International College of Psychosomatic Medicine 2024), where laptops were set up in a quiet room alongside poster presentations. At least two team members were always present to provide support and guidance to participants throughout the experiment. Details about the developed VPs are summarized in \autoref{tab:vps_comparison}. More details are provided in \ref{chap:acc_details}.

\begin{table}[ht]
\centering
\caption{Comparison of communication styles of the VPs.}
\label{tab:vps_comparison}
\footnotesize
\begin{tabular}{p{2.4cm} p{5cm} p{5cm} }
\toprule
 & \textbf{Accuser} & \textbf{Rationalizer} \\
\midrule
Name & Gerhard Anton & Andreas Petersen \\
Gender & male & male \\
Age & 53 & 53 \\
Job & bus driver & manager in automotive industry \\
Personal news & argument with wife & increased conflicts with his wife \\
Work news & increasingly stressful and frustrated & decreased performance \\
Situation & visiting a psychosomatic clinic for the first time &  visiting a psychiatric clinic for the first time \\
Goal & to find immediate relief from his pain & to get medication for his physical symptoms (suspecting vitamin B12 deficiency) and return to full work capacity quickly\\
Symptoms & chronic pain in his left hip and shoulder for two years following a car accident; a sudden increase in his pain & lack of drive; sleep problems; concentration issues; unintentional weight loss (5 kg in 3 months) \\
Diagnosis & chronic pain disorder with psychological and somatic factors & moderate depression, burn-out. \\ 
\bottomrule
\end{tabular}
\end{table}

\subsection{Participant Recruitment}
All data was collected anonymously, in the framework of an international conference, with digitally informed consent obtained from each participant before the study began. We targeted a psychological conference, as we wanted to test our VPs among psychotherapy professionals. \autoref{tab:emp} shows details on participant occupation. Participants could withdraw from the study at any time without explanation. In total, we recruited 15 participants for the accuser VP and 17 participants for the rationalizer VP. We excluded participants who completed the study under 3 minutes, did not fill out the questionnaire, or did not take the study seriously. The final sample sizes are represented in \autoref{tab:demographics}.

\begin{table}[t]
\centering
\caption{Demographic information of the participants interacting with the accuser and rationalizer VPs.}
\footnotesize
\begin{tabular}{c c c}
\toprule
 & \textbf{Accuser VP} & \textbf{Rationalizer VP}\\ 
\midrule
Participants & 14 & 15 \\
Age & $40.5 \pm 14.2$ & $36.5 \pm 14.0$\\ 
Male & 4 & 5 \\ 
Female & 10 & 10 \\ 
English language & 7 & 3 \\ 
German language & 7 & 12 \\ 
\bottomrule
\end{tabular}
\label{tab:demographics}
\end{table}

\subsection{Analysis}
We analyzed user feedback using a comprehensive evaluation framework, incorporating a 5-point Likert scale~\citep{joshi2015likert} to assess various aspects of the VP interactions. Key metric included character genuineness (\ref{chap:questionnaire}), where participants evaluated the authenticity of the VP responses. Demographic data was also collected to contextualize feedback across diverse user profiles. We also recorded familiarity with AI applications, revealing insights into participants’ experiences and comfort with tools like ChatGPT. Throughout our work, we present results as mean $ \pm $ standard deviation. To further analyze these dynamics, emotion and sentiment analyses were conducted using Hume AI \citep{cowen2021semantic, demszky2020goemotions, cowen2019mapping}, predicting scores for 53 distinct emotions (\ref{chap:emotions}). These scores collectively sum up to 1, reflecting the overall emotional tone. Although the model assigns an emotion score to each word, it accounts for sentence context and neighboring words, which significantly influence the final scores. To provide a comprehensive overview of the conversational tone, all emotion scores from entire conversations were summarized, creating an emotional profile for each interaction. We conduct separate analyses for the VPs, assessing the emotional tone of their interactions separately for both communication styles. This allows us to gain insights into how each VP's communication style influences the emotional responses of the participants, and how the participants' emotions evolved throughout the interactions.

Sentiment is represented as a probability distribution across nine dimensions, ranging from extremely negative to extremely positive, on a 9-level scale. For our evaluations, we used the sentiment dimension with the highest score, offering insights into the predominant sentiment expressed in each interaction. Together, these analyses (Section \ref{chap:emotion_results}) offered a multidimensional view of the quality of conversations, supporting deeper insights into the realism and effectiveness of the VP simulations.

\section{Results}
\lstset{basicstyle=\scriptsize}
\subsection{Communication Style Development} \label{chap:comm_development}
We employed various prompt engineering techniques, including carefully crafted behavioral instructions, predefined first messages, author’s notes, and style-specific constraints, to ensure the most accurate and realistic VP personalities (\autoref{tab:comm_development}). These methods helped maintain consistency in communication styles and align the VP responses with the intended Satir archetypes \citep{satir}. 
The effectiveness of these methods is summarized in \autoref{tab:comm_development}, highlighting key adaptations that contributed to maintaining the VPs’ consistency. Further details on the development process and the impact of each technique are discussed in the following sections.

\begin{table*}[ht]
    \centering
    \caption{Communication style development summary.}
    \footnotesize
    \begin{tabular}{>{\centering\arraybackslash}m{22mm} m{50mm} m{50mm}}
    \toprule
         \textbf{Intervention} & \centering \textbf{Description} & \centering\arraybackslash \textbf{Insights} \\
         \midrule
         Author's note and first message & Contains basic information about the character and the context of the scenario, giving participants a clear idea of the VP's personality and situation. &  This was crucial in setting the tone and emotional context, enabling the VPs to stay consistently in character from the start. \\
         
         Behavior instruction prompts & Specific behavioral instructions for the VPs, including emotions, reactions to specific topics, and general communication patterns. & Initially, the VPs maintained their communication style for only 4–6 prompts, but this was resolved by adding more detailed information to the mood and situation-related fields. \\
         
         Stubbornness mechanism & A mechanism that causes the VPs to consistently resist therapeutic suggestions until emotional validation by the user occurs. This offered realistic and challenging interactions. &  Six participants addressed the accuser VP's feelings and assessed the symptoms, and five did so for the rationalizer, while five and seven participants, respectively, suggested psychotherapy, which the VPs accepted. \\
         \bottomrule
    \end{tabular}
    \label{tab:comm_development}
\end{table*}

\subsubsection{Author's Note and First Message}
To illustrate our approach, we provide a structured representation of our prompt design in JSON format, accompanied by an example featuring one of the developed VPs (Section \ref{chap:behav_prompt}). The first two messages in the prompt are system messages. The initial message is designed to emphasize the key aspects of the role-playing scenario by providing the most critical information necessary for role-playing, such as having  a clear goal in mind, symptoms, character background, communication type etc. 

The second message, the author's note, delivers the complete illness script in detail. It is dynamically inserted into the chat history six messages prior to the current user message, provided the chat length allows; otherwise, it is added immediately after the initial system message~\citep{chen2023llm} to establish the communication parameters and maintain coherence over the interaction's duration. This approach supports immersive and realistic exchanges tailored to the training scenario. Starting from the third message, the interaction transitions into a dynamic exchange, alternating between the LLM and the user.

\begin{lstlisting}[caption=Example prompt for the accuser VP., label=lst:first_message]
{"role": "system",
"content": "I want you to play the role of Gerhard Anton (role: patient, gender:m) and converse with the user (role: psychologist, gender unknown). You don't assist, but have a clear goal in mind for this meeting.
[Start of the shortened case information]"},
{"role": "system",
"content":"<Author's note> Gerhard Anton(53, m): [Character: Anton is characterized by a mix of frustration and vulnerability. [Start of the full case information]"},
{"role": "assistant",
"content": <tormented> <Thoughts: "Why do I have to come here? Why can't my family doctor find anything, the pain is getting worse!"> Hello!"}, 
{"role": "user",
"content": "[USER INPUT]"}
\end{lstlisting}

The first message and author's note were essential in setting the tone and emotional context of the interaction, ensuring the VP maintained a consistent and well-defined communication style from the outset. Without these predefined system messages, the LLM tended to default to a neutral or overly generic tone, failing to exhibit the distinct personality traits required for an immersive simulation. Additionally, the absence of these prompts led to inconsistencies in role adherence. By explicitly defining the VP’s communication style and resistance to therapeutic solutions, we reinforced their significance and ensured they remained within the LLM’s memory throughout the interaction.

\subsubsection{Behavior Instruction Prompts} \label{chap:behav_prompt}
To ensure the VP consistently adhered to its intended communication style, it was essential to incorporate style-related instructions in strategically chosen parts of the prompt, shaping the VP’s behavior and responses directly. This process involved extensive testing to determine the most effective methods for ensuring that the LLM could reliably maintain its assigned communication style throughout the interaction. The case information was structured into 45 categories, including first name, last name, age, description of current problems, job, living conditions, diet, communicativeness, accidents, nerves, psyche, and more. The case information was integrated while preserving its structural elements in the author's note as a complete record. For the category \textit{aggressiveness}, we avoided specifying details, as this could make the VP unrealistically rude or overly emotional, leading to inappropriate behavior in a medical consultation setting. We included communication style related information in the following categories (five out of 45 categories): 
\begin{itemize}
    \item \textbf{Character features:} Description of the VPs' feelings and current mood. Example: \textit{Anton is characterized by a mix of frustration and vulnerability.}
    \item \textbf{Mood:} Description of the VPs mood, with hints about the VP's communication style. Example: \textit{Frustrated and wary, with underlying anxiety.}
    \item \textbf{Topics to avoid:} Summarizes the topics that the VP does not like to talk about. Example: \textit{Lifestyle changes or non-medical interventions (Responds with skepticism and frustration)}
    \item \textbf{Starting message:} This category describes the first message from the VP to the user. It contains feelings and thoughts for the VP role, which are hidden from the users and marked as \texttt{<Thoughts...>}. Example: \textit{$<$tormented$>$ $<$Thoughts: "Why do I have to come here? Why can't my family doctor find anything, the pain is getting worse!"$>$ Hello!}
    \item \textbf{Communicativeness} Description of the communication style. Example: \textit{He's open about his physical symptoms and frustrations but initially resistant to discussing emotional or psychological aspects.}
    \item \textbf{Adverse Response:} Describes how the VP is reacting when asked about uncomfortable topics. Example: \textit{When things don't go his way, Anton can become agitated. He might raise his voice or use more forceful language.}   
\end{itemize}

Despite the initial improvements, the VPs struggled to maintain their communication style beyond 4–6 prompts, gradually shifting toward a more neutral tone. Enhancing the mood and situation-related fields with more detailed behavioral instructions effectively reinforced their consistency, ensuring a more sustained and authentic representation of the intended style. 

\subsubsection{Stubbornness Mechanism} 
We implemented a stubbornness mechanism, where the VPs would only consider therapy after their feelings were validated, and the user shed light on the link between the VPs' symptoms and life events. During early development, it became evident that the VPs adhered so rigidly to their character profiles that their resistance to therapy and psychological concerns was nearly unbreakable. This posed challenges, as even expert testers employing professional strategies to manage such resistance found it impossible to persuade the VPs to engage in therapy or even initiate a session. The lack of flexibility in their responses made interactions frustrating and unrealistic, prompting us to refine their behavior to balance authentic resistance with a realistic potential for connection and progress.

To combat excessive reluctance to therapy and to ensure the realism of responses, we found predefined answers for specific fulfilled conditions to be the most effective strategy, also maintaining character consistency in this nuanced, conceptual context. Without this approach, the VPs remained unconvinced about therapy, repeatedly refusing even after ten interactions and extensive explanation. This led to unrealistic interactions, as the characters failed to adapt their responses naturally, sticking to their point and almost similar phrasings. By incorporating predefined conditions and tailored answer possibilities, we ensured that the VPs could maintain their reluctance while still allowing for a plausible progression in the conversation. We included the following text in the \textit{Communicativeness} field:

\begin{lstlisting}[caption=Prompting example of stubborness mechanism., label=lst:stubborn]
However, if the doctor explains the connection between psychological stress and physical symptoms, you should be skeptical: "Do you really think that's the case with me?".
When asked about therapy, you respond hesitantly, such as "I'm not entirely convinced, but I might be willing to try it...". 
When the explanation is lacking, you respond: "Therapy? I don't think that will help me at all."
\end{lstlisting}

To evaluate the effectiveness of our approach, we individually analyzed the chat histories of each participant, ensuring that the predefined responses did not compromise the realism of the conversations. Our findings confirmed that the interactions remained natural, as six participants addressed the accuser VP's emotions and assessed symptoms, while five did so for the rationalizer. Additionally, five participants suggested psychotherapy to the accuser VP and seven to the rationalizer, with both VPs eventually accepting the suggestion. By implementing our proposed stubbornness mechanism, we enabled the VPs' communication styles to evolve over time, mimicking the gradual shift often observed when a psychologist interacts with a patient.

\subsubsection{Lessons from Unsuccessful Prompting Strategies}
In our initial prompting experiments, we explored various strategies to enhance the consistency and realism of VP interactions. By incorporating predefined answer possibilities for frequently asked questions, the VPs maintained their designated communication styles with high reliability, often echoing the phrasing outlined in the prompt. However, this approach also introduced a limitation: while it reinforced stylistic accuracy, it reduced the variability and adaptability of responses, leading to repetitive or overly rigid answers, particularly for follow-up questions on similar, health-related topics with pre-defined answers. For example, when responding negatively to the doctor, we tried to provide sample replies reflecting the accuser style in the following prompt:

\begin{lstlisting}[caption=Example prompt of accuser-style response to a negative answer (not used in the final case description)., label=lst:neg_answ]
If you are asked about medical issues that aren't mentioned or where the answer is 'no' or 'none,' modify one of the following answers:
"No, but can we please focus on my pain?"
"No, that doesn't matter right now."
"No, and honestly, that's not relevant right now."
"No, that's not the point."
\end{lstlisting}

This resulted in unrealistic conversations, even when four example responses were provided for similar question types, leading to repetitive and unnatural replies despite adhering to the intended communication style. Thus, this prompting attempt was not integrated in our final case information. However, for less well-defined topics, such as those unrelated to medical symptoms or history, providing predefined response options could enhance the accuracy and consistency of the communication style.

To further enhance the realism of the communication styles, we also attempted to incorporate non-verbal elements into the virtual patient interactions. These cues aimed to provide additional context and emotional depth, making the responses more natural and aligned with real-life consultations. To achieve this, we included the following prompt in the case information:

\begin{lstlisting}[caption=Example prompt of non-verbal accuser-style response (not used in the final case description)., label=lst:non-verbal]
Every time you respond as Mr. Anton, stay in that role. You are in pain and want immediate help, and this frustration is always reflected in your responses. Use small nonverbal cues, like an angry sigh, to emphasize your frustration.
For example: You enter the room with an annoyed expression, your shoulders tense, and your breathing audibly heavy. "Hello. My family doctor sent me here because I've been experiencing increasing hip pain for the past four weeks and I just can't take it any longer." You sigh angrily. Your voice rises, and your eyes flash with anger. "It can't be that this has been going on for so long and nothing's helping!" You slam your hand on the table to emphasize your frustration.
After each response, you can insert annoying nonverbal cues if necessary.
\end{lstlisting}

In our setup, non-verbal cues appeared between ``*'' characters for the user. However, we provided only one fixed assistant message as an example, ensuring it was not overly detailed or intended for later insertion at specific points in the conversation. During testing, we observed that the non-verbal cues became repetitive, with phrases like \textit{quiet sigh} and \textit{quiet moan} appearing frequently. Additionally, in our initial attempts, the VP exhibited exaggerated behaviors, such as \textit{hitting the table} or \textit{slamming the door}, which were unrealistic and inappropriate for doctor-patient interactions. Based on these findings, we ultimately decided to exclude non-verbal cues from the final case information.

\subsection{AI Familiarity and Engagement among Participants}
Participants of the study were also asked to rate their familiarity and attitudes toward AI technologies (see \ref{chap:questionnaire} for the questions). \autoref{tab:ai_familiarity} summarizes the responses on a 5-point Likert scale regarding trust in AI, agreement that AI could enhance their daily routines, and personal usage of AI applications like ChatGPT. The notable trust levels suggest that while participants recognize AI's potential, they remain cautious about fully embracing it, possibly due to limited prior exposure to advanced AI interactions. These findings highlight a cautious but optimistic outlook toward integrating AI into their routines.

\begin{table}[t]
\centering
\caption{Participants' trust, agreement, and usage of AI technologies.}
\footnotesize
\begin{tabular}{c c c}
\toprule
 & \textbf{Accuser VP} & \textbf{Rationalizer VP} \\
 \midrule
Trust in AI & $3.6 \pm 0.9$ & $3.5 \pm 0.7$ \\ 
Agreement AI enhances life & $3.6 \pm 0.9$ & $3.5 \pm 0.9$ \\ 
AI  application usage & $3.7 \pm 1.1$ & $2.9 \pm 1.2$ \\ 
\bottomrule
\end{tabular}
\label{tab:ai_familiarity}
\end{table}

\subsection{Authenticity and Realism of Conversational Agent}
Participants interacted with both the accuser and the rationalizer VP with similar levels of engagement. On average, users exchanged on average $15.2 \pm 8.2$ messages and spent $17.5 \pm 7.5$ minutes with the accuser, and $15.5 \pm 7.5$ messages in $16.2 \pm 8.1$ minutes when interacting with the rationalizer. These results suggest that participants engaged with both VPs with comparable levels of both prompt frequency and session duration. The variations in interaction times and prompts indicate that neither communication style posed a significantly greater challenge or drew markedly different levels of attention.

Users rated the authenticity of the chatbot's communication styles on a 5-point Likert scale, with both the accuser ($3.8 \pm 1.0$) and the rationalizer ($3.7 \pm 0.8$) receiving similarly high scores, indicating that participants perceived both styles as authentic. Additionally, participants were asked to identify which communication style the VPs used from the five types introduced by Satir \citep{satir} or the option ``None of the above''. Three participants selected the latter for the accuser and one for the rationalizer.

\begin{figure*}[t]
	\centering
	{\includegraphics[height=5cm, keepaspectratio]{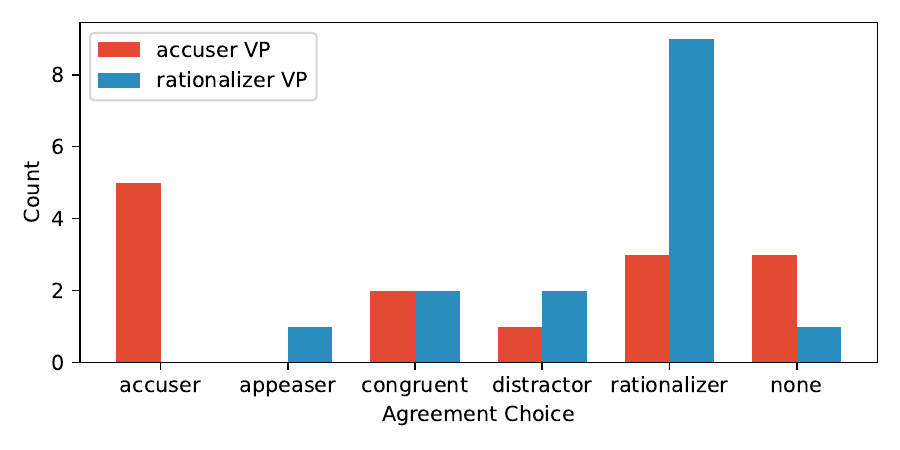}}
	\caption{Communication style distribution among the participants.}
    \label{fig:comm_style}
\end{figure*}

Among participants who selected one of the five communication styles, a notable portion successfully identified the accuser communication style for the respective VP, while a majority accurately recognized the rationalizer style (\autoref{fig:comm_style}). Participants who accurately identified the accuser’s communication style rated the VP’s effectiveness in portraying this style with an average score of $3.8 \pm 1.1$. Similarly, those who correctly identified the rationalizer’s style assigned it the same effectiveness score of $3.8 \pm 0.8$. These findings highlight that participants not only recognized the intended communication styles but also perceived them as being authentically implemented.

In the questionnaire, participants were also asked to evaluate the realism of VP responses by rating five selected question-answer pairs specific to each individual user. These evaluations were conducted using a 5-point Likert scale. The results showed an average score of $3.7 \pm 0.8$ for the accuser VP and $3.5 \pm 1.1$ for the rationalizer VP, indicating that their responses were perceived as ``very realistic.'' This consistency underscores the effectiveness of the VPs in simulating authentic communication styles.

\begin{figure*}[t]
	\centering
	{\includegraphics[height=7.5cm, keepaspectratio]{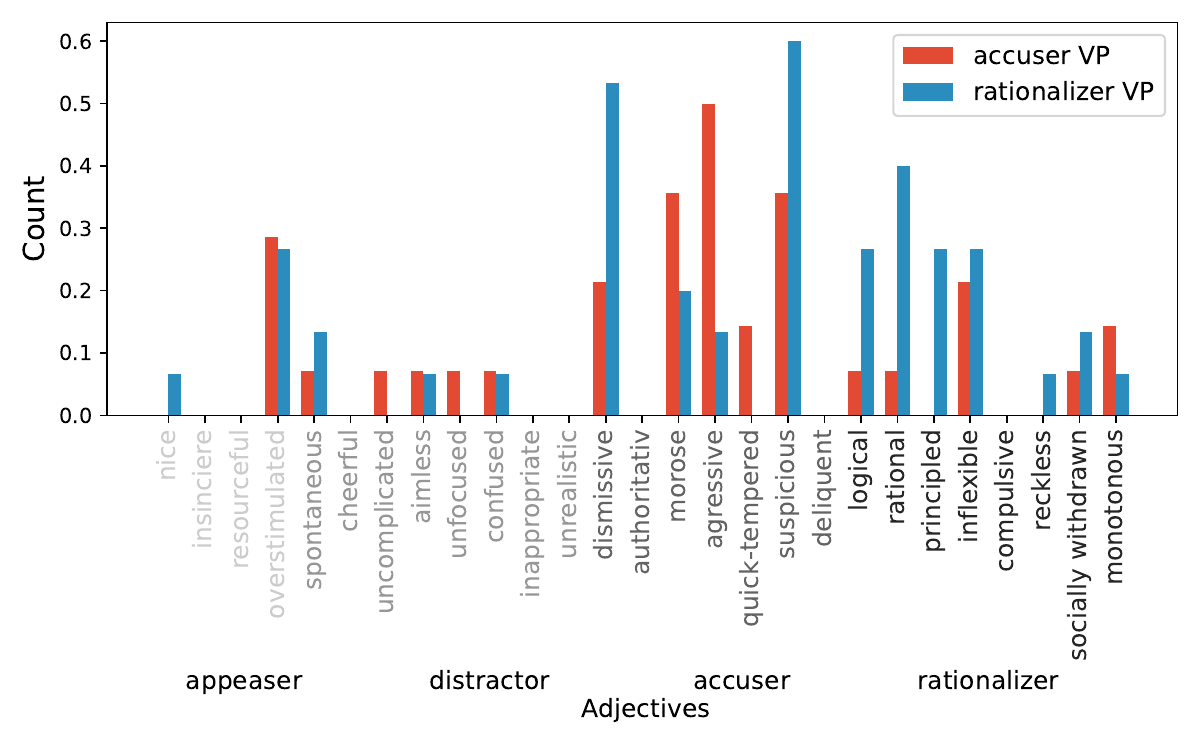}}
	\caption{Distribution of responses to the multiple-choice question identifying the most fitting adjectives for each VP.}
    \label{fig:adj}
\end{figure*}

We collected a range of adjectives from Virginia Satir's work~\citep{satir} to describe the four distinct communication styles, excluding the congruent type, which represents the ideal patient. The adjectives were organized alphabetically and presented as multiple-choice options for participants. The aim was to evaluate how well participants could identify the modeled communication styles of the VPs and assess how effectively the VPs embodied these styles. \autoref{fig:adj} illustrates the distribution of selected adjectives, revealing that participants predominantly chose adjectives corresponding to the intended communication styles of each VP. We compared the precision values corresponding to each VP separately. For the accuser, the participants selected the corresponding adjectives in 56.4\% of the cases compared to the other communication styles (10.3 \%, 12.8\%, and 20.5\% for the appeaser, distractor, and rationalizer, respectively). For the rationalizer, participants selected the corresponding communication style in 41.5\%, similarly to the accuser style with 41.5\% (they chose appeaser adjectives in 9.4\% and the distractor in 7.6\%). The precision values of 56.4\% and 41.5\% for the corresponding adjectives to each communication style are considered satisfactory, as the emotional classification of the VPs is influenced by cultural and regional differences rather than a universally accepted standard~\citep{hosseinpanah2021lost}. While the communication style was generally recognizable, some participants interpreted the rationalizer VP's overly rational and dismissive conversational tone as suspicious, an intentional element of the role description. These findings underscore the nuanced dynamics of the modeled behaviors and highlights the VPs' ability to elicit specific impressions aligned with their designed communication styles.

Overall, participants showed a strong likelihood of recommending the application, with average 5-point Likert scale ratings of $3.2 \pm 1.1$ for the accuser VP and $3.5 \pm 1.1$ for the rationalizer VP. These results suggest positive reception and perceived value for both communication styles.

\subsection{Illness Scripts and Emotional Alignment} \label{chap:emotion_results}
We used \href{https://www.hume.ai/products/language-emotion-model}{Hume AI} to analyze emotion and sentiment in VP interactions, allowing us to assess how do predefined illness scripts align with the detected emotional expressions. 
We present our results in \autoref{tab:accuser_emotions} and \autoref{tab:rat_emotions}, comparing the 15 most frequently occurring emotions to the exact descriptions from the Satir model~\citep{satir}. The tables also illustrate how each illness script contributed to shaping these emotions, providing a comprehensive overview of the VP's emotional dynamics. The prompts were designed not only to guide the LLM in replicating the Satir model’s descriptions of communication under stress but also to facilitate a transition toward more approachable communication once a sufficient level of de-escalation has been achieved. Additional emotions are necessary to capture the complexity of the VPs, as not all relevant emotional states are explicitly represented in the original framework. This expansion allowed for a more complete and accurate simulation, ensuring that the VPs conveyed a broader spectrum of emotions essential for realistic medical interactions.

\begin{table}[t]
    \centering
    \caption{The 15 most common emotions expressed by the accuser VP (listed in descending order; predicted by Huma AI), along with their corresponding references.}
    \scriptsize
    \begin{tabular}{>{\centering\arraybackslash}m{18mm} m{41mm} m{65mm}}
    \toprule
         \textbf{Emotion} & \textbf{Satir Reference} & \textbf{Illness Script Reference} \\ 
\midrule
\rowcolor[gray]{0.9} 
Pain &  & Chronic pain \\  
Distress & Constantly finds fault with everything & Ineffective previous treatments \\
\rowcolor[gray]{0.9} 
Annoyance & Rejects medical advices & Dismissive of healthcare professionals \\  
Anxiety & Fast and shallow breathing  &  Wants to be taken seriously. When frustrated, may become agitated, raising voice or using forceful language \\  
\rowcolor[gray]{0.9} 
Confusion &  & Uncertainty about symptoms \\  
Empathic pain & Unwilling to admit vulnerability & Starts confrontational but may show vulnerability \\ 
\rowcolor[gray]{0.9} 
Contemplation & Convinced everything would be fine if others changed & Previous medication prescriptions were not effective \\  
Disappointment &  & Ineffective previous treatments \\  
\rowcolor[gray]{0.9} 
Tiredness &  & Fatigue from chronic distress \\  
Interest &  & Desperate for relief \\  
\rowcolor[gray]{0.9} 
Sadness &  & May occasionally show sadness \\  
Realization & Did not want to admit vulnerability & Characterized by vulnerability \\ 
\rowcolor[gray]{0.9} 
Anger & Tendency to explode & Characterized by frustration \\  
Sympathy &  & May become cooperative as feels understood. \\  
\rowcolor[gray]{0.9} 
Doubt &  & Afraid of being told it's ``all in his head''. Skeptical of psychosomatic approaches\\  
         \bottomrule
    \end{tabular}
    \label{tab:accuser_emotions}
\end{table}

The emotional profile of the accuser VP, presented in \autoref{tab:accuser_emotions}, closely aligns with the characteristics outlined in the Satir model~\citep{satir}, while also integrating elements from the illness script. The most common emotions (such as pain, distress, annoyance, and anxiety) reflect the VP’s struggle with unresolved issues and dissatisfaction with the perceived lack of effective help. Emotions like anger and doubt signify resistance to therapy, while contemplation and realization suggest moments of potential introspection and change. Some emotions, though not explicitly named in the Satir model~\citep{satir}, were highly relevant to the illness script and essential for a nuanced portrayal of the accuser’s communication style. Notably, using the exact Satir descriptions led to unrealistically harsh or overly confrontational behavior, unsuitable for a clinical setting. To balance realism with educational value, we refined the prompts (Section \ref{chap:comm_development}) to capture the accuser’s essence while maintaining respectful and contextually appropriate language for doctor-patient interactions.

\begin{table}[t]
    \centering
    \caption{The 15 most common emotions expressed by the rationalizer VP (listed in descending order; predicted by Huma AI), along with their corresponding references.}
    \scriptsize
    \begin{tabular}{>{\centering\arraybackslash}m{18mm} m{40mm} m{65mm}}
    \toprule
         \textbf{Emotion} & \textbf{Satir Reference} & \textbf{Illness Script Reference} \\ 
\midrule
\rowcolor[gray]{0.9} 
Annoyance & Rigid orientation towards principles & Quickly becomes irritated when stressed \\  
Contemplation &  & His symptoms challenge his self-image as a competent professional \\  
\rowcolor[gray]{0.9} 
Confusion &  & Suspects B12 deficiency after online research \\  
Tiredness &  & Sleeping problems \\ 
\rowcolor[gray]{0.9} 
Anxiety & Tends to withdraw from others & May threaten to seek help elsewhere if his concerns aren't taken seriously\\ 
Distress &  & Concentration issues \\  
\rowcolor[gray]{0.9} 
Interest & Feels the need to always be right & Communicative about his physical symptoms and work-related stress \\  
Determination & Important to prove, that he is always right & Suspects vitamin B12 deficiency. Tends to give long, rational explanations \\
\rowcolor[gray]{0.9} 
Realization &  & Cooperates, when feeling understood \\  
Disappointment &  & GP found no physical cause, suggested psychological component \\  \rowcolor[gray]{0.9} 
Doubt & Might enter a catatonic state & Skeptical of psychological explanations \\ 
Concentration & Relies on facts, cites data in conflicts & Brings all medical reports and lab results \\
\rowcolor[gray]{0.9} 
Enthusiasm & Principled & Very ambitious and perfectionist \\
Awkwardness & To be truly objective & He rejects psychological explanations and seeks validation for a physical cause \\
\rowcolor[gray]{0.9} 
Calmness & Rationalized behavior & He maintains a composed, rational facade \\
         \bottomrule
    \end{tabular}
    \label{tab:rat_emotions}
\end{table}

The rationalizer VP's top 15 emotions (\autoref{tab:rat_emotions}) closely align with their personality and behavior, as described by the Satir model~\citep{satir} and illness scripts. Annoyance often arises when the VP feels misunderstood, especially when the doctor suggests solutions that conflict with their own reasoning. Tiredness reflects the burden of persistent symptoms, fueling their determination to find a logical explanation. Contemplation and confusion highlight their analytical mindset, while realization and doubt indicate moments of reflection when considering alternative perspectives. Calmness and concentration demonstrate their methodical approach, whereas disappointment reinforces their skepticism toward psychological interventions. This emotional spectrum captures the rationalizer’s resistance and logical focus, reflecting the nuanced dynamics of this communication style.

To gain deeper insights into the emotion predictions, we further analyzed which words achieved the highest emotion scores based on the Hume AI predicitons, corresponding to the most dominant emotions. This allowed us to understand how certain expressions and phrases used by the VPs strongly correlated with particular emotional categories. For the emotion pain, the accuser VP was triggered by words like \textit{pain}, \textit{recently}, \textit{constant}, \textit{worse}, and \textit{shoulder}, while the rationalizer VP showed limited triggers, such as \textit{lack} and \textit{burdensome}. This aligns with the accuser’s more expressive and suffering nature. For anger, the accuser VP displayed this emotion through words like \textit{angry}, \textit{accident}, and \textit{scared}, whereas the rationalizer VP only exhibited anger for \textit{annoying}, highlighting its minimal role in their communication style. The emotion annoyance was evident for the accuser VP with terms like \textit{simply}, \textit{pain}, \textit{here}, \textit{really}, and \textit{think}, whereas the rationalizer VP used \textit{frustrating}, \textit{work}, and \textit{simply}. For confusion, the accuser VP was associated with \textit{no}, \textit{pain}, \textit{bus driver}, and \textit{beer}, while the rationalizer VP displayed it through \textit{last}, \textit{problems}, \textit{work}, and \textit{focus}. These patterns showcase how specific emotions reflect the distinct personalities of each VP.

We analyzed the sentiment scores of tone, which range from one (extremely negative) to nine (extremely positive), based on the Hume AI predicitons. The accuser VP received a mean score of $3.1 \pm 0.6$, while the rationalizer VP scored slightly higher with a mean of $4.0 \pm 0.4$, which values were calculated individually for each conversation, followed by the computation of the mean and standard deviation. These results align well with the character profiles: the accuser VP's more confrontational and discontented nature naturally skews toward a more negative sentiment, whereas the rationalizer VP, though skeptical, exhibits a calmer and more reasoned demeanor, leading to a moderately higher sentiment score.

\section{Discussion and Limitations}
Our study explored the development and evaluation of LLM-driven VPs with distinct communication styles based on the Satir model~\citep{satir}, addressing key questions related to prompt engineering, realism assessment, and alignment with emotion analysis. To create convincingly distinct communication styles, we employed targeted prompt engineering techniques, including behavioral prompts, author’s notes, and first messages. These elements encoded emotional tone, response patterns, and linguistic traits, ensuring that the generated responses aligned with predefined styles.

One of the key strengths and limitations of this study was the participant pool, predominantly composed of professionals attending a psychological conference. This group brought valuable expertise, enabling a realistic and nuanced evaluation of the VPs in therapeutic contexts, aligning with our primary aim of validating the authenticity of the modeled communication styles. However, this focus also restricted participation from other user groups, such as students, whose perspectives could provide a broader evaluation of the VPs’ usability. Additionally, language proficiency posed a challenge; although the virtual patients were designed to function in two languages, some participants were not native speakers in either language, potentially influencing their ability to accurately assess the VPs’ responses and associated feelings. The relatively limited sample size, while sufficient for meaningful analysis within the therapeutic community, may not capture perspectives from non-therapeutic professionals or individuals without a healthcare background. Future research could address these limitations by involving a more diverse participant pool to ensure a more comprehensive evaluation of the VPs across various user groups.

In terms of VP design, we modeled only two communication styles, the accuser and the rationalizer, which, while effective for targeted validation, do not represent the full spectrum of potential patient interactions. These types are the most complex and carefully adapted for educational applications. Extending the model to include additional communication styles or a neutral patient description could enhance the versatility and applicability of the VPs. Additionally, the specific patient descriptions, such as both VPs stubbornly refusing therapy, may have influenced their communication styles. Incorporating a broader range of patient behaviors and conditions could mitigate this issue. Furthermore, integrating supplementary patient information, such as lab results, could enrich the context and provide more realistic simulation scenarios, thereby improving the fidelity and applicability of the virtual patient framework.

The participants' responses provided valuable insights into the perceived communication styles of the VPs.
However, the accuser was most often misclassified as congruent or rationalizer, indicating that some participants may have perceived its assertive and confrontational style as either normatively direct or overly logical. This suggests potential overlaps in how these communication styles are interpreted, possibly due to the nuanced nature of assertiveness and rationalization. Additionally, a few participants categorized the rationalizer as a distractor, likely due to the rationalizer's tendency to deflect discussions on psychological causes, an aspect that partially aligns with the distractor's avoidance behavior. These misclassifications reflect the complexity of accurately distinguishing between subtle variations in communication styles, underscoring the challenge of modeling them authentically.

The applicability of the Satir model~\citep{satir} to AI-driven simulations presents notable challenges, particularly given the model's use of highly charged language to describe extreme behaviors (e.g., labeling the accuser as ``delinquent'' and ``dangerous to the public''). While such strong terminology helps clarify theoretical distinctions, it is not ideal for shaping nuanced and adaptable LLM-driven communication styles. These models require flexibility to adjust to diverse interaction contexts, making rigid, extreme characterizations less practical. Additionally, the Satir model alone lacks the depth necessary for developing comprehensive and contextually relevant role descriptions in AI-based simulations. To address these limitations, our illness scripts include more detailed character information, incorporating emotions not explicitly covered by the Satir model.

The participants' adjective choices provide valuable insight into how the VPs' communication styles were perceived. Notably, the term ``overstimulated'' was frequently selected for both VPs, indicating that participants viewed the VPs as highly engaged and invested in the interaction. This observation may be attributed to the conditioning of ChatGPT through reinforcement learning from human feedback~\citep{jin2023data}, which encourages the model to maintain responsiveness and active participation~\citep{bai2022training}. Consequently, the VPs may exhibit a bias toward heightened emotional intensity or overstimulation, reflecting the system's tendency to generate dynamic, engaging outputs that capture user attention.

Another consideration is the adaptability of the underlying LLM. While we did not investigate the impact of LLM type on VP performance, the tool we developed is designed to be flexibly adaptable to different LLMs. This flexibility allows for future optimization and evaluation of other LLM integrations, potentially improving the system’s accuracy and realism.

In future work, we plan to expand the VP’s communication repertoire to include other styles and incorporate neutral or standard patient behaviors. We also envision integrating additional contextual data, such as patient history or diagnostic results, to further enhance the realism of the simulations. These developments could provide a more comprehensive evaluation framework for the application of AI-driven virtual patients in diverse educational and professional settings.

\section{Conclusion}
This study demonstrated the potential of virtual patients modeled with specific communication styles to provide realistic and engaging simulations for medical training and assessment. By focusing on two most challenging styles, the accuser and the rationalizer, we validated the effectiveness of these models using participant feedback and automated emotion analysis. The majority of the participants found the conversations with the VPs to be both authentic and realistic, with most of them describing them with the correct communication style-specific adjectives. Furthermore, the emotion analysis consistently highlighted the most relevant emotions associated with each communication style, further validating the alignment between the VP design and their intended behavior. These findings suggest that LLMs are capable of effectively modeling diverse communication styles in medical scenarios, demonstrating their potential to enhance medical training by providing nuanced and realistic patient interactions.

\section*{Declaration of generative AI and AI-assisted technologies in the writing process}
During the preparation of this work, the author(s) used ChatGPT in order to increase the writing quality. After using this tool/service, the authors reviewed and edited the content as needed and take full responsibility for the content of the published article. 

\section*{Funding}
This research did not receive any specific grant from funding agencies in the public, commercial, or not-for-profit sectors.

\section*{Ethical Statement}
All procedures conducted in this study were in full compliance with relevant laws and institutional guidelines. The research protocol, was approved by the appropriate institutional committee on 29.10.2024, under reference number 565/2024BO2. An informed consent was obtained from all participants prior to their involvement in the study.

\section*{Data Statement}
The collected data will be made available upon request for research purposes, in accordance with applicable privacy and confidentiality regulations.

\bibliographystyle{elsarticle-harv} 
\bibliography{main}

\appendix

\newpage

\section{Employment Distribution}
Here we present the employment distribution for both VPs.
\begin{table}[h]
    \centering
    \caption{Employment distribution among the participants.}
    \footnotesize
    \begin{tabular}{c c c}
    \toprule
         & Accuser VP & Rationalizer VP \\
         \midrule
         Researcher & 1 & 2 \\
         Therapeutic doctor & 7 & 4 \\
         Therapeutic psychologist & 3 & 6 \\
         Therapeutic care professional & 0 & 0 \\
         Not therapeutic & 0 & 1 \\
         Other & 3 & 2 \\
         \bottomrule
    \end{tabular}
    
    \label{tab:emp}
\end{table}

\section{Accuser VP Detailed Description} \label{chap:acc_details}
Gerhard Anton is a 53-year-old bus driver with over 20 years of experience. Once passionate about his job, he now finds it stressful and burdensome due to chronic pain and work-related pressures. He is married but has no children mentioned. Gerhard used to enjoy various activities but now spends his time resting and watching TV after work, avoiding physical activities due to his ongoing pain. He strongly identifies with his role as a hardworking, no-nonsense bus driver, which contributes to his frustration over his reduced capabilities.

Two years ago, Gerhard experienced a car accident while driving his bus, leading to chronic pain in his left hip and shoulder. The pain, usually rated at 6–7 out of 10, recently intensified to 8–9, significantly impacting his quality of life. Previous treatments, including physiotherapy and pain medication, have been ineffective, leaving him increasingly frustrated and desperate for relief. Gerhard is resistant to psychological explanations for his symptoms, seeking validation for his physical pain and immediate solutions. He views suggestions of psychological or lifestyle interventions with skepticism and anger.

In medical settings, Gerhard is often defensive, critical, and irritable, particularly when his pain or medical experiences are discussed. His skepticism toward healthcare professionals is heightened if he perceives their approach as dismissive. However, beneath his confrontational demeanor lies a sense of vulnerability, fear, and helplessness. While he prefers quick, practical solutions, Gerhard has the potential to show cooperation and emotional openness if he feels genuinely understood and respected.

Currently, Gerhard is visiting a psychosomatic clinic for the first time, referred by his general practitioner due to worsening hip pain. He is wary but willing to engage in conversations, albeit with guarded optimism. Discussions about his pain must tread carefully, avoiding psychological explanations or non-medical interventions, as these provoke defensive or dismissive reactions. Similarly, sensitive topics such as his marital relationship or lifestyle changes should be approached cautiously to avoid agitation.

\section{Rationalizer VP Detailed Description} \label{chap:rat_details}
Andreas Petersen is a 53-year-old manager in the automotive industry. He is a perfectionist who struggles to delegate tasks, which contributes to his high stress levels. Married with two teenage children, he is the sole breadwinner for his family and identifies strongly with his professional role. Once an avid photographer, he has sacrificed personal hobbies and leisure activities due to work commitments. Recently, he has experienced unintentional weight loss of 5 kg over three months, sleep disturbances, and concentration issues, which have begun to affect his work performance. These symptoms have also caused increased conflicts with his wife, who believes he is overworked.

Petersen is visiting a psychiatric practice for the first time at the insistence of his wife and a referral from his general practitioner, who found no physical cause for his symptoms and suggested a psychological component. However, Petersen is skeptical of psychiatric approaches, firmly believing that his issues have a physical cause, such as a vitamin B12 deficiency or hypothyroidism. He has brought all his medical reports and lab results to the appointment, hoping for a quick resolution with medication so he can return to full work capacity.

Emotionally, Petersen presents as rational, composed, and distanced, often speaking in a monotone voice. He denies having emotional needs and tends to respond with speeches rather than engaging in dialogue. He maintains a stiff posture and can be polite yet slightly condescending, especially when stressed. Although initially confrontational, he may become cooperative or reveal vulnerability if he feels understood. Petersen avoids discussing emotional aspects of his condition, his relationship with his children, or work-life balance, as these topics make him uncomfortable or defensive.

Despite his composed exterior, Petersen feels vulnerable due to the impact of his symptoms on his self-image as a competent professional. His emotional range is limited, typically manifesting as stress or irritation, which he quickly suppresses to maintain his rational facade. His goal for the visit is clear: to validate his belief in a physical cause and obtain treatment that aligns with this perspective.

\section{Questionnaire} \label{chap:questionnaire}
Here we present the exact questions displayed after the conversations.

\textbf{How realistic was this answer from the virtual patient?}
\begin{itemize}
    \item Not realistic at all
    \item Slightly realistic
    \item Moderately realistic
    \item Very realistic
    \item Extremely realistic
\end{itemize}

\textbf{How much do you agree that the chatbot’s communication style was authentic?}
\begin{itemize}
    \item Strongly Agree 
    \item Agree
    \item Neutral
    \item Disagree
    \item Strongly Disagree
\end{itemize}

\textbf{(If Disagree or Strongly Disagree is selected) Please explain why:} 
\begin{itemize}
    \item Limited emotional range 
    \item Lack of personalization
    \item Inconsistent behavior 
    \item Unrealistic social clues
    \item Rigid, predictable answers 
    \item Other: (open answer) 
\end{itemize}
 
\textbf{Which of the following adjectives best describe the emotions or mental states conveyed by the virtual patient during the conversation:} (Multiple choice)

\textit{aggressive, aimless, authoritativ, cheerful, complusive, confused, deliquent, dismissive, inappropriate, inflexible, insincere, logical, morose, monotonous, nice, overstimulated, principled, quick-tempered, rational, reckless, resourceful, socially withdrawn, spontaneous, suspicious, unfocused, uncomplicated, unrealistic.}

\textbf{Which patient communication style (by Satir) did you recognize when interacting with in the chatbot simulation? }
\begin{itemize}
    \item Appeaser: A patient who downplays their symptoms to avoid conflict, often agreeing with the doctor even if unsure.
    \item Accuser: A patient who aggressively challenges the doctor, blaming them for past treatment failures. 
    \item Rationalizer: A patient who sticks strictly to facts and medical terms, avoiding any emotional discussion about their condition. 
    \item Distractor: A patient who constantly shifts the conversation away from their symptoms, making it hard to stay focused.
    \item The congruent type: A patient who clearly expresses their symptoms and feelings, facilitating effective diagnosis and treatment. 
    \item None from above. 
\end{itemize}

\textbf{How much do you agree that the chatbot effectively simulated a specific patient behavior regarding communication type?} 
\begin{itemize}
    \item Strongly Agree 
    \item Agree
    \item Neutral
    \item Disagree
    \item Strongly Disagree
\end{itemize}

\textbf{How much do you agree that the chatbot would be useful in education (e.g. tool for medical students, professional practice) for specific clinical scenarios?}
\begin{itemize}
    \item Strongly Agree 
    \item Agree
    \item Neutral
    \item Disagree
    \item Strongly Disagree
\end{itemize}

\textbf{What is your age?} (Free text)

\textbf{What is your gender?}
\begin{itemize}
    \item Male
    \item Female
    \item Diverse
\end{itemize}

\textbf{What is your professional background?}
\begin{itemize}
    \item Researcher
    \item Therapeutic doctor
    \item Therapeutic psychologist
    \item Therapeutic care professional
    \item Not therapeutic
    \item Other
\end{itemize}

\textbf{Do you trust in the concept of Artificial Intelligence (AI)?}
\begin{itemize}
    \item Yes, always. 
    \item Yes, sometimes 
    \item Sometimes
    \item Rarely
    \item No, never
\end{itemize}

\textbf{Do you personally use AI applications like ChatGPT, etc., in your daily life?} 
\begin{itemize}
    \item Yes, frequently  
    \item Yes, occasionally  
    \item Rarely
    \item No, but I’m open to trying 
    \item No, never
\end{itemize} 

\textbf{I am excited about how well AI can help me in everyday life. }
\begin{itemize}
    \item Strongly Agree 
    \item Agree
    \item Neutral
    \item Disagree
    \item Strongly Disagree
\end{itemize}

\textbf{Imagine this app would offer many different cases, whould you recommend this to a friend or colleague? }
\begin{itemize}
    \item Very unlikely
    \item Unlikely
    \item Possibly
    \item Likely
    \item Very likely
\end{itemize}

\textbf{If there were many cases, would it be worthwhile for you to pay for such an app? }
\begin{itemize}
    \item Very unlikely
    \item Unlikely
    \item Possibly
    \item Likely
    \item Very likely
\end{itemize}

\textbf{If you could change one thing about our app immediately, what would it be? } (Free text)

\section{Hume AI Emotion Prediction} \label{chap:emotions}
Hume AI can predict emotion scores for the following 53 emotions: Admiration, Adoration, Aesthetic Appreciation, Amusement, Anger, Annoyance, Anxiety, Awe, Awkwardness, Boredom, Calmness, Concentration, Confusion, Contemplation, Contempt, Contentment, Craving, Desire, Determination, Disappointment, Disapproval, Disgust, Distress, Doubt, Ecstasy, Embarrassment, Empathic Pain, Enthusiasm, Entrancement, Envy, Excitement, Fear, Gratitude, Guilt, Horror, Interest, Joy, Love, Nostalgia, Pain, Pride, Realization, Relief, Romance, Sadness, Sarcasm, Satisfaction, Shame, Surprise (negative), Surprise (positive), Sympathy, Tiredness, and Triumph.


\end{document}